\documentclass[twocolumn,showpacs,aps,prl]{revtex4}
\input{epsf}

\newcommand{\gsim}{\raisebox{-0.7ex}{$\;\stackrel{\textstyle >}{\textstyle\sim}\;$}}

\begin{document}
\title{High-frequency blockade in a tight-binding one-dimensional lattice
with single vibrating atomic state}

\author{M.V. Entin}
\email{entin@isp.nsc.ru}
\affiliation{Institute of Semiconductor
Physics, Siberian Division, Russian Academy of Sciences,
\\Novosibirsk, 630090 Russia}

\author{M.M. Mahmoodian}
\email{mahmood@isp.nsc.ru}
\affiliation{Institute of Semiconductor
Physics, Siberian Division, Russian Academy of Sciences,
\\Novosibirsk, 630090 Russia}
\altaffiliation[Also at ]{Novosibirsk State University,
Novosibirsk, 630090, Russia}

\begin{abstract}
One-dimensional tight-binding lattice, single site of which
possesses harmonically vibrating level is studied. The states of
non-interacting electrons incident with fixed energy from infinity
are considered. It is shown that at definite conditions the site
reflects electrons {\it absolutely and elastically}
(high-frequency blockade states). The problem is treated both
numerically and (in the case of narrow band) analytically. The
results are compared with the free-electron 1D problem with
vibrating $\delta$-functional potential. Together with the
blockade states the local and reflectionless states are examined.
Possible realization of the system as a lattice of quantum dots
is discussed.
\end{abstract}

\pacs{73.50.Pz, 73.23.-b, 85.35.Be} \maketitle

The transport through systems with local vibrating potential
attracted attention in connection to the quantum pumps
\cite{Thouless,Brouwer,Switkes,Kim1,Moskalets02,Moskalets03,Zhu,Kim,brag,mah,mah1}.
One interesting feature of these  systems  is the presence of
states with no transmission (high-frequency blockade states)
\cite{Bagwell,Wagner,Wagner1,Reichl,Reichl1,Tkach1,Tkach2,mah2,mah3}.
The blockade states are unusual and impossible in conservative
systems. They give a lot of new possibilities of control of
electron transport in nanostructures. In particular, the absolute
reflection by a single vibrating well leads \cite{mah2,mah3} in
double-well structure to appearance of local states situated on
the background of continuum. These states produce very narrow
controlled resonances in the system transparency.

Here we concentrate on the problem of electron states in the
tight-binding model with the vibrating level of a single site.
This problem permits both numerical and, in the case of narrow
permitted band, analytical consideration. We shall study the
problems of  continuous (scattering) and discrete eigenstates. The
model permits coexistence of such discrete states as blockade,
reflectionless and local states which will be the topic of our
special attention. From this standpoint, the tight-binding model
is much reacher than the considered earlier models. In addition,
we shall discuss the possibility of practical realization of
systems with local vibrating potential.

We study a 1D tight-binding lattice expressed by a system of
equations
\begin{equation}\label{tight}
i\dot{a}_m-\delta_{m,0}(u+v\cos(\omega
t))a_0+\frac{\Delta}{2}(a_{m+1}+a_{m-1})=0.
\end{equation}
The Eq. (\ref{tight}) corresponds to a lattice with  the
overlapping amplitudes between sites $\Delta/2$, the energy is
referred to the levels of sites with $n\neq 0$. We use the
quantities $u,~v,~\Delta,~E$ measured in units of frequency
$\omega$; $\hbar=1$.

In absence of the vibrating site ($u=v=0$) the eigenstates of the
system $a_m=e^{ipm-iE(p)t}$ with quasimomentum $p$ has the energy
$E(p)=-\Delta \cos p$. The propagating electron can possess energy
within the permitted band $-\Delta<E<\Delta$.

If $v=0, ~~u\neq 0$ the site $n=0$ represents an ''impurity''
which produces scattering of propagating electrons and the
impurity state above or below the permitted band. In presence of
vibrations ($v\neq 0$) the energy does not conserve changing by
some quanta $\omega$.

We consider two kinds of states, namely, the scattering problem
with given energy of an incident electron and local quasienergy
(Floquet) states; the latter are permitted in the tight-binding
model, unlike the free-electron model with a vibrating
delta-potential.

The general solution of the scattering problem reads
\begin{eqnarray}\label{tightwf1}
a_{m\leq 0}=\sum_n\left(\delta_{n,0}e^{ip_0m}+r_ne^{-ip_nm}\right)e^{-i(E+n)t},\nonumber\\
a_{m\geq 0}=\sum_n t_ne^{ip_nm}e^{-i(E+n)t}.
\end{eqnarray}
Here $r_n=t_n-\delta_{n,0}$, $p_n$ is the solution of the equation
$E+n=-\Delta \cos p_n$ corresponding to the positive velocity of
the states inside the permitted band, in accordance with
causality. Some of states get to the forbidden band; for them the
positive imaginary value of $p_n$ should be chosen to guarantee
the decay of states apart from the zero site.

The transition amplitudes $t_n$ satisfy an equation
\begin{eqnarray}\label{sys}
t_n(\Lambda(E+n)-u)-\frac{v}{2}(t_{n+1}+t_{n-1})=\Lambda(E)\delta_{n,0}.\end{eqnarray}
Here $\Lambda(E)=i\sqrt{\Delta^2-(E+i\cdot0)^2}$ (the term
$i\cdot0$ reflects the causality). In general, the system
(\ref{sys}) can be solved numerically.

The system under consideration can be treated as a quantum wire
constructed from a long chain of quantum dots the ends of which
are connected to electron seas (source and drain).  Suppose, that
the source and the drain of the device have equilibrium  electron
gas at zero temperature. The chemical potentials of seas are
different due to the applied dc voltage. The solutions of the
scattering problem determine the system conductance:
\begin{equation}\label{G}
G=G_0\int dE\sum_n|t_n|^2\left(-\frac{\partial f(E)}{\partial
E}\right).
\end{equation} Here $f(E)$ is the Fermi distribution function;
$G_0=2e^2/h$ is the conductance quantum.

The presence of mentioned blockade states reflects in the
vanishing of the conductance at definite conditions.

\section*{Numerical results}

\begin{figure}[ht]
\centerline{\epsfysize=5cm\epsfbox{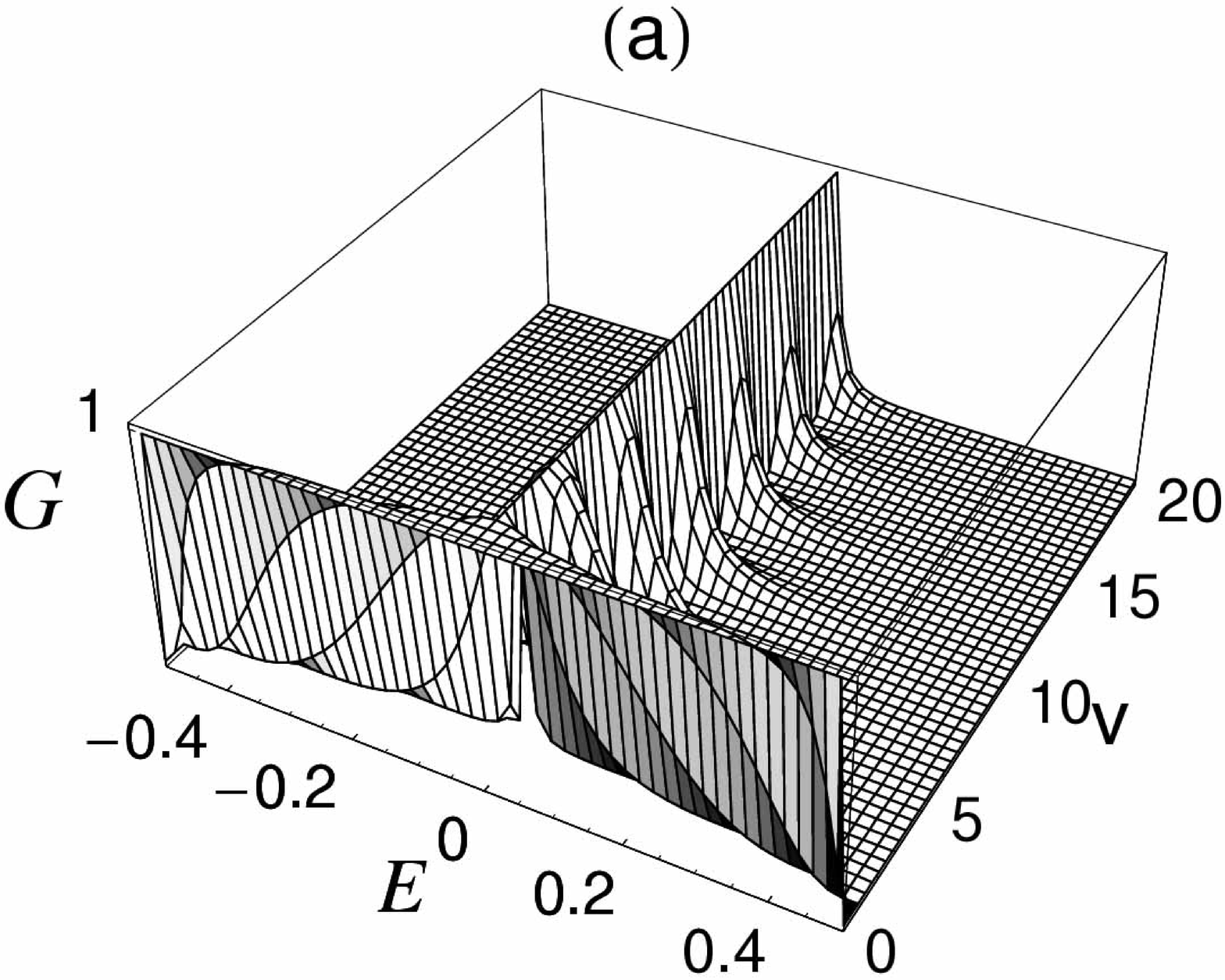}}
\centerline{\epsfysize=5cm\epsfbox{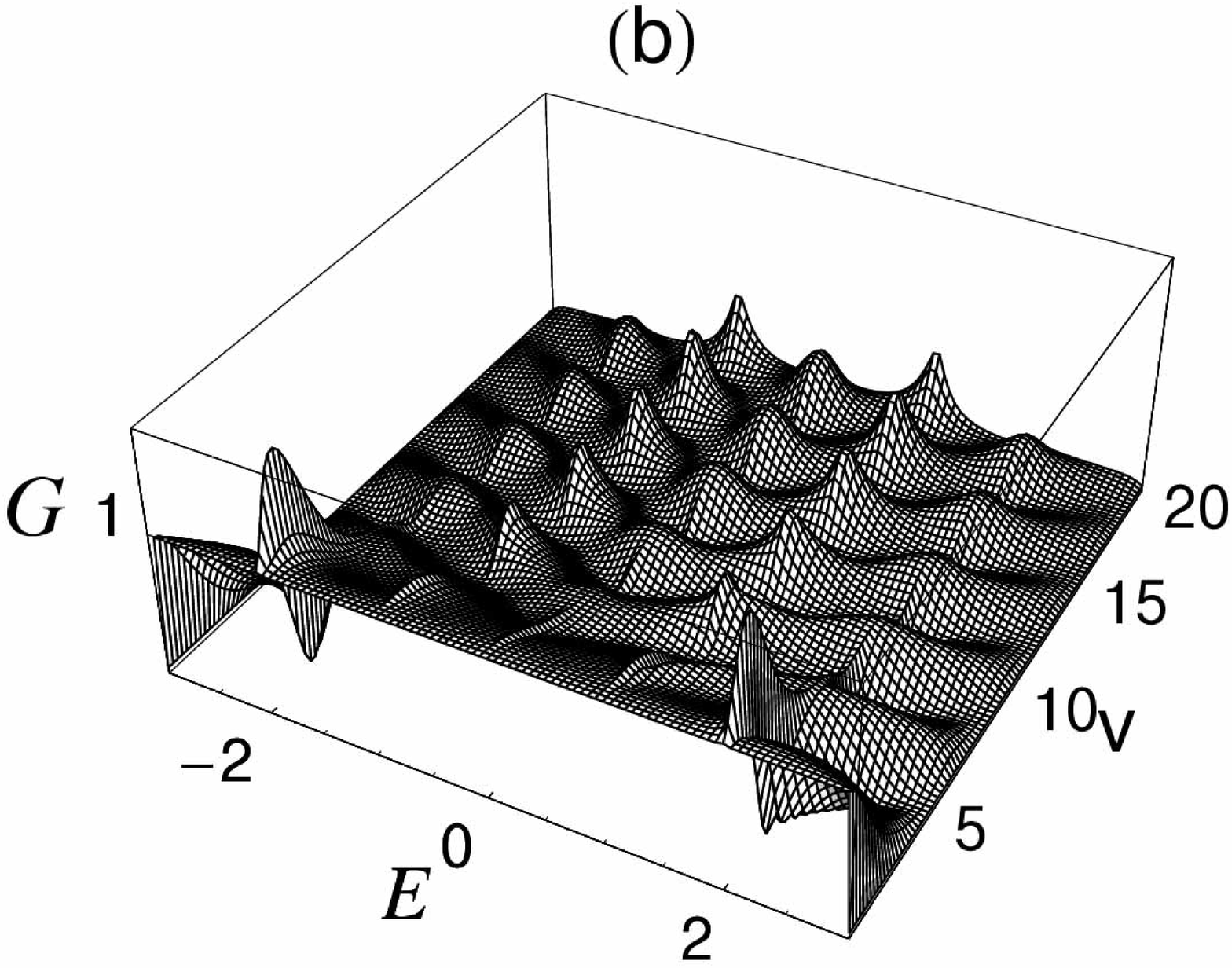}}
\caption{Conductance $G$ (in units of $G_0$) versus the Fermi
energy and parameter $v$ at $u=0$, $\Delta=0.5$ (a) and
$\Delta=2.5$ (b). The Fig. (a), where $\Delta<1$, distinctly shows
the reflectionless state $E=0$, $G=1$. For $\Delta>1$ (Fig. (b))
the reflectionless state disappear.}\label{gev}
\end{figure}

\begin{figure}[ht]
\centerline{\epsfysize=6cm\epsfbox{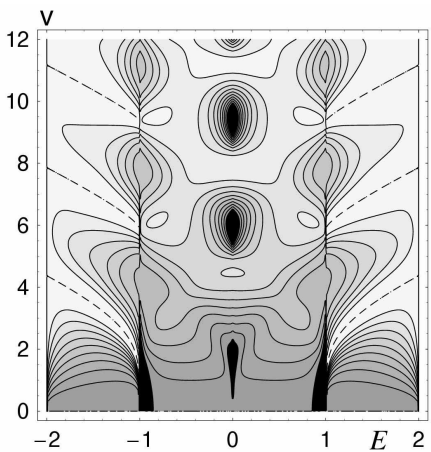}}\caption{Map of
conductance levels (in units of $G_0$) as a function of the Fermi
energy and the parameter $v$ at $u=0$, $\Delta=2$. The levels run
equidistantly from $<10^{-6}$ (white) to $\geq 1$ (black). The
blockade states are depicted by dashed curves.}\label{bl}
\end{figure}

\begin{figure}[ht]
\centerline{\epsfysize=6cm\epsfbox{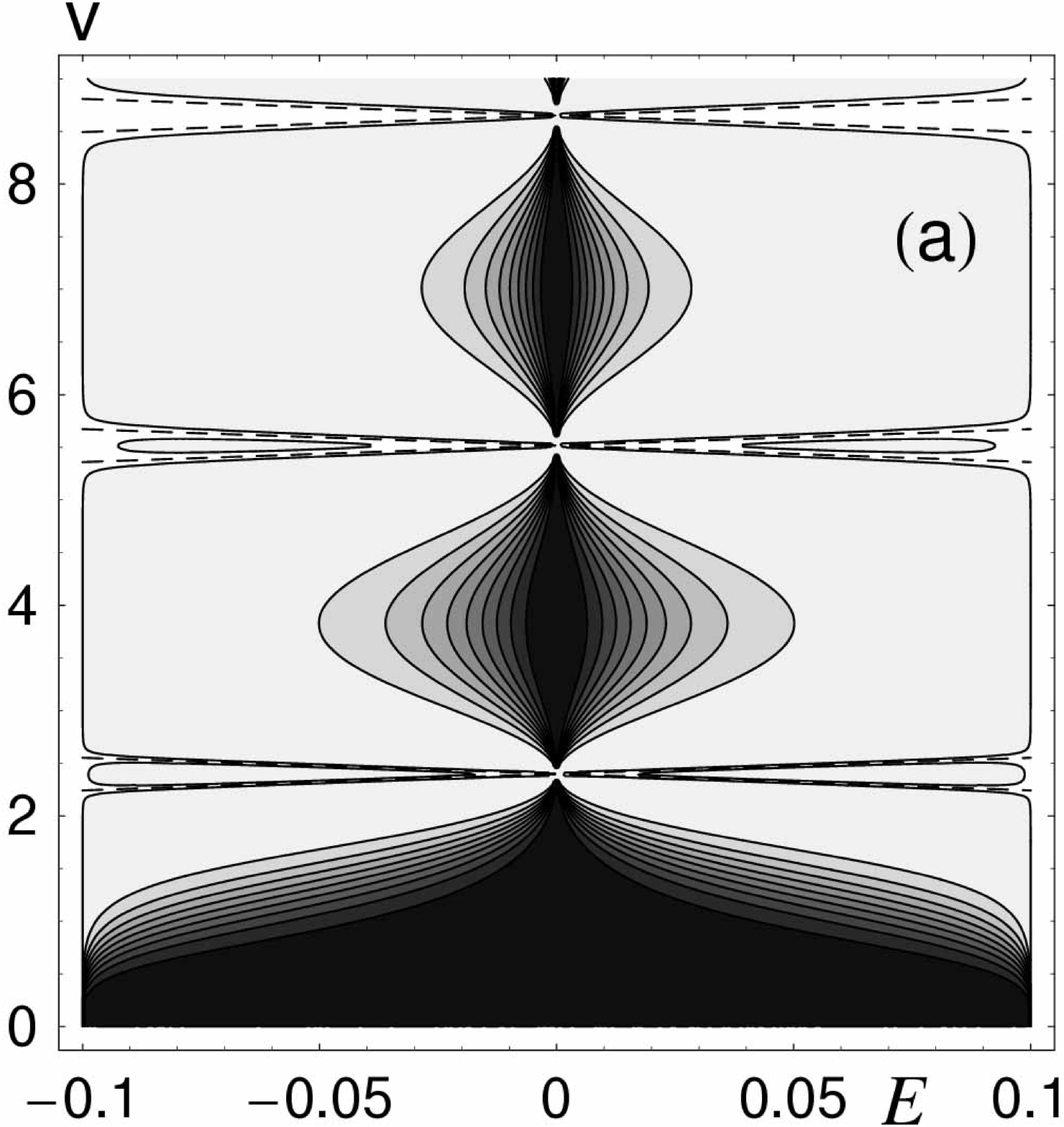}}
\centerline{\epsfysize=6cm\epsfbox{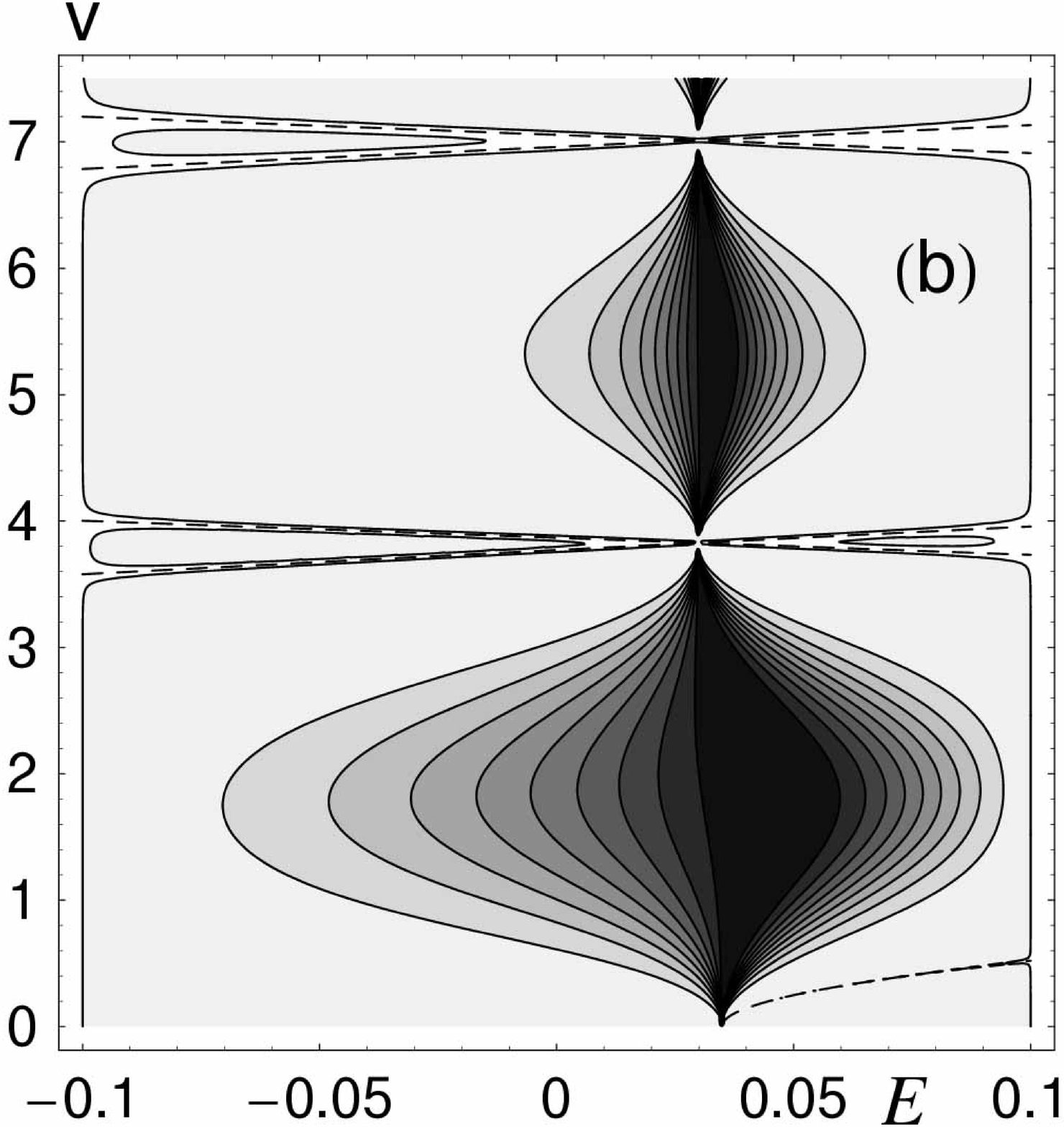}}\caption{The same as
in Fig. \ref{bl} at $\Delta=0.1$ and $u=0$ (a), $u=0.13$ (b). In
the last case the picture loses symmetry $E\leftrightarrow -E$.
Blockade states cross all permitted band.}\label{bl1}
\end{figure}

\begin{figure}[ht]
\centerline{\epsfysize=5.5cm\epsfbox{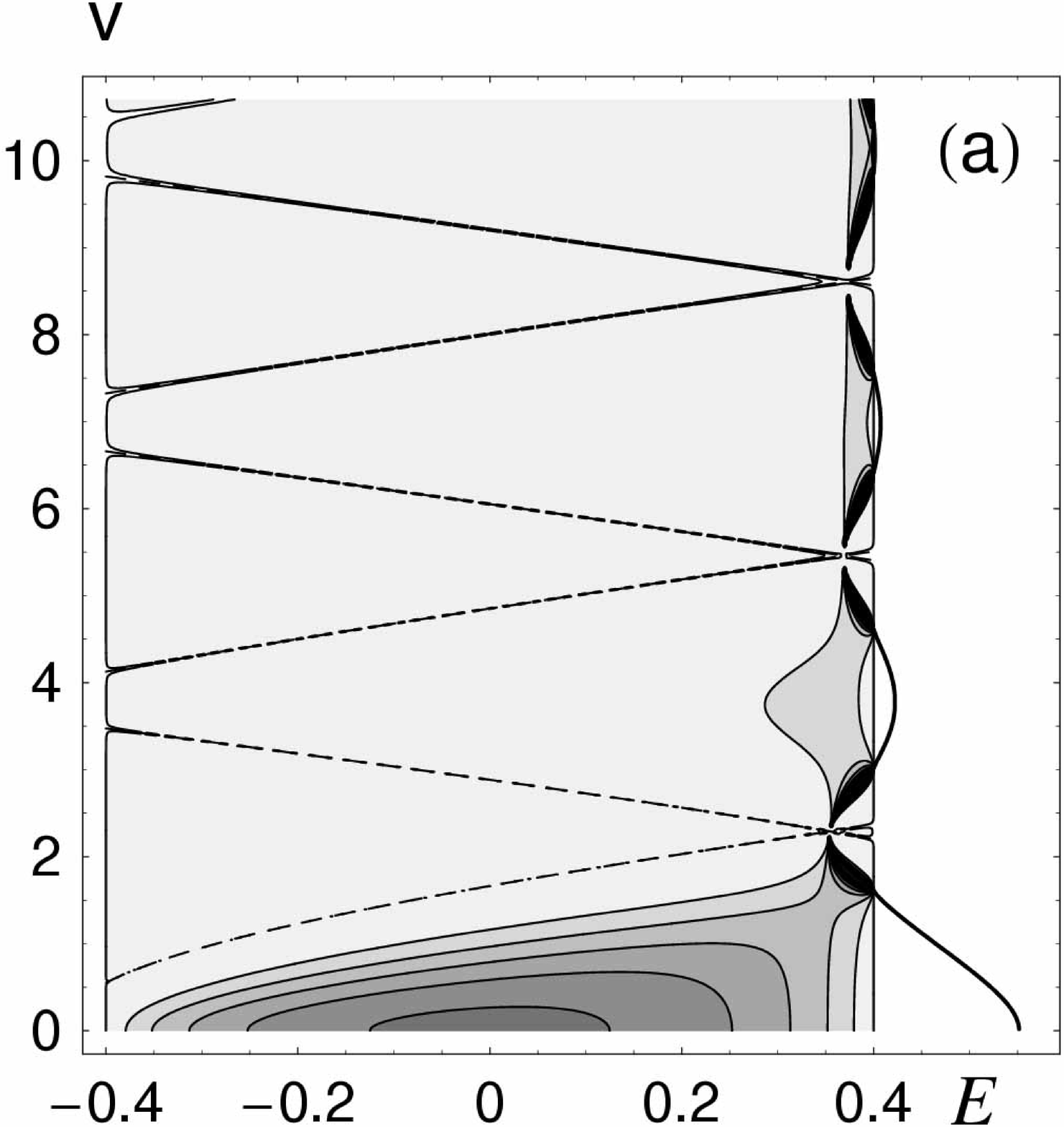}\epsfysize=5.5cm\epsfbox{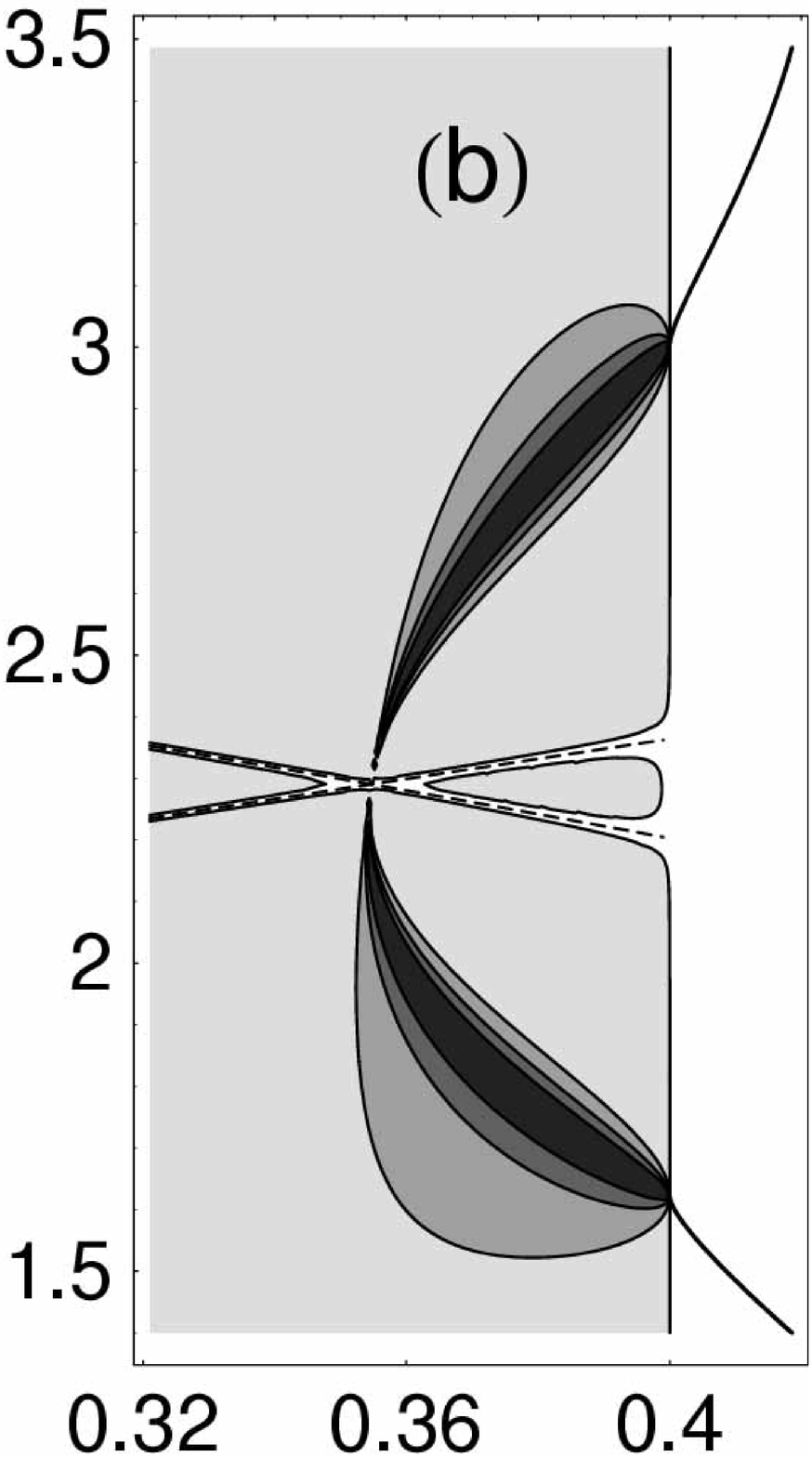}}
\caption{(a): The same as in Fig. \ref{bl} at $\Delta=0.4$,
$u=0.38$. In this case the crossing point of the blockade and
reflectionless states approach the permitted band boundary
$E=0.4$. The local states (solid lines) are situated to the right
of this boundary. (b): Magnified neighborhood of the first
crossing point of blockade states and reflectionless states
illustrates how the local states convert to the reflectionless
states (inside black regions).}\label{bl-loc2}
\end{figure}

The calculated conductance at zero temperature is depicted in the
Figs. \ref{gev}-\ref{bl-loc2} for different values of $\Delta$.
The Fermi energy, denoted below as $E$, runs from the bottom
$-\Delta$ to the top $\Delta$ of the permitted band.

The conductance oscillates with $v$, that reflects the
interference nature of the process. At $u=0$ the pictures are
symmetric with respect to the band center (Figs. \ref{gev},
\ref{bl}). This electron-hole symmetry can be easily deduced from
the Eq. (\ref{sys}).  For large $\Delta>1$ (Figs. \ref{gev}b,
\ref{bl}) the picture exhibits  singularities, connected with the
thresholds $\pm \Delta$, namely $\pm (\Delta-n)$. The
singularities $n=0,1$ are most pronounced. The other singularities
(easing with $n$) arise as photon repetitions of these
singularities. For a wide permitted band $\Delta\gg 1$ the
behavior of the conductance near the band boundaries  is similar
to the model of free electrons interacting with the vibrating
$\delta$-potential \cite{mah2,mah3}. This is not the case for a
narrow permitted band.

There are two other important features of the conductance
dependence. One is the vanishing of the conductance at specific
energies $E(u,v)$ (blockade states). It means that at
corresponding energy all channels of transmission become closed,
$t_n=0$ for $|E+n|<\Delta$.

These high-frequency blockade states can be found directly. Note
that if for some $n>0$ $t_n=t_{n+1}=0$, then all $t_k=0$ for
$k>n+1$. Hence, the condition $t_0=t_1=0$ automatically solves all
equations (\ref{sys}) with $n>0$. The equations for $n<0$ compose
a homogeneous system, while the equation for $n=0$ determines the
value for $t_{-1}$. This condition can be fulfilled if
$-\Delta+1>E>-\Delta$. If the energy satisfies the condition
$\Delta-1<E<\Delta$, above mentioned is valid with the change of
sign in front $n$. The homogeneous system has solutions for some
eigenvalues of parameters. Practically, they can be found as
eigenvalues for $1/v$, if one rewrites the system as
\begin{eqnarray}\label{sys2}
\frac{1}{v}t_n-\frac{1}{2(\Lambda(E+n)-u)}(t_{n+1}+t_{n-1})=0,~~~~~
n<0.\end{eqnarray}

The Figs. \ref{bl}, \ref{bl1}a show the dependence of blockade
states on parameters for $u=0$. If $\Delta>1$ (Fig. \ref{bl}) the
domains of existence of the blockade states tight to the left and
right band boundaries do not overlap, if $\Delta<1$ (Fig.
\ref{bl1}a) the blockade states cross, and if $\Delta<1/2$ they
overlap all permitted band.

Another feature is the existence of a state with absolute
transmission when the conductance exactly equals to 1. For $u=0$
this reflectionless state corresponds to the center of permitted
band $E=0$. It exists if $\Delta<1$. It is interesting that there
are essentially singular points where the blockade and
reflectionless states coexist.

For $u\neq 0$ the relief of the conductance becomes asymmetric
and, in particular,  the line of the reflectionless state moves
from the permitted band center $E=0$ and curved.

For finite $u$ besides the considered states the local states
appear in the forbidden band. These states are constructed from
the waves decaying apart from the site $n=0$. They obey the Eq.
(\ref{tightwf1}) with no incident wave for energy which lies in
the forbidden band (all $p_n$ are chosen with positive imaginary
part). These states originate from the local impurity state at
$v=0$ when the transitions into propagating modes caused by
alternating field are forbidden by the energy conservation law.
Accordingly, their quasienergy plus (or minus) any integer number
should not get into the permitted band.  The local states appear
below or above the forbidden band in dependence on the sign of
$u$.

Similar to resonance tunneling in conservative systems, the local
states transform to the reflectionless states when they go through
the permitted band boundaries $E=\pm \Delta$ (Fig. \ref{bl-loc2}).

\section*{Analytical results}
Let us consider the limit $\Delta\ll 1$. For states with energy
$E\gg \Delta$ in the zero approximation one can replace
$\Lambda(E)$ by $E$, and if $E\ll 1$ rewrite the system
(\ref{sys}) as
\begin{eqnarray}\label{sys1}
\left(H-\lambda\right)_{n,n'}t_{n'}=\Lambda(E)\delta_{n,0},
\end{eqnarray} where $\lambda=u-E$ and $H=H^{(0)}+V$,
\begin{eqnarray}\label{ham0}
H_{n,n'}^{(0)}=n\delta_{n,n'}-\frac{v}{2}(\delta_{n,n'-1}+\delta_{n,n'+1}).
\end{eqnarray}

The perturbation $V$ reads as
\begin{eqnarray}\label{pert0}
V_{n,n'}=(\Lambda(E+n)-E-n)\delta_{n,n'}.
\end{eqnarray}

In the limit of small $\Delta$ all elements of $V_{n,n'}$ are
negligible except for $n=0$, and
\begin{eqnarray}\label{pert}
V_{n,n'}=(\Lambda(E)-E)\delta_{n,0}\delta_{n',0}.
\end{eqnarray}

The Eq. (\ref{sys1}) with local perturbation Eq. (\ref{pert}) can
be solved using the Green function. The eigenvectors $\psi_k$ of
the Hamiltonian $H^{(0)}$ are given by the integer Bessel
functions $\psi_k=J_{n-k}(v)$ whose eigenvalues are $k$. The Green
function $\hat{G}^{(0)}=1/\hat{H}^{(0)}$ is
$$G_{n,n'}^{(0)}=\sum_k\frac{J_{n-k}(v)J_{n'-k}(v)}{k-\lambda}.$$
Using the Green function, we find
\begin{eqnarray}\label{trans0}
t_n=\frac{\Lambda(E)\sum\limits_k\frac{J_{n-k}(v)J_{-k}(v)}{k-\lambda}}
{1+[\Lambda(E)-E]\sum\limits_k\frac{J_k^2(v)}{k-\lambda}}.
\end{eqnarray}
In the considered case only the channel $n=0$ belongs to the
propagating states. Hence the zeros of $t_0$ determine the
blockade states, while the poles do the localized states:
\begin{eqnarray}
\sum\limits_k\frac{J_k^2(v)}{k-\lambda}=0 ~~~~~\mbox{for
blockade states,}\label{states1}\\
1+[\Lambda(E)-E]\sum\limits_k\frac{J_k^2(v)}{k-\lambda}=0~~~~~\mbox{for
local states.}\label{states2}
\end{eqnarray}

\noindent{\it Blockade states.} If $u\ll 1$ then the terms $k=0$
with the denominator $\lambda=u-E$ majorizes the sums, and Eq.
(\ref{states1}) gives the energies of the blockade states. For
$v\ll 1$ one can keep two terms of the series in Eq.
(\ref{states1}) with $k=0$ and $k=N$. As a result we obtain
\begin{eqnarray}\label{block0}
E_0=\tilde{u}+\frac{N}{(N!)^2}\left(\frac{v}{2}\right)^{2N}
\end{eqnarray}

The blockade states also exist near the points $v=v_{0,m}$, where
$v_{n,m}$ are the zeros of the Bessel functions $J_n(v)$. If $u$
is close to some integer $N$: $u=N+\tilde{u}$, $\tilde{u}\ll 1$ we
obtain
\begin{eqnarray}\label{block1}
E_m=\tilde{u}\pm\frac{J_N(v)}{\Big[\sum\limits_{k\neq
N}J_k^2(v_{N,m})/(k-N)^2\Big]^{\frac12}.}
\end{eqnarray}

\noindent{\it Local states.} The local states satisfy to an
equation
\begin{eqnarray}\label{loc}
u=N+E\left[1-\left(1-\sqrt{1-\Delta^2/E^2}\right)J_N^2(v)\right].
\end{eqnarray}

If the alternating signal $v=0$, the local states convert to the
static impurity states $E=u\sqrt{1+\Delta^2/u^2}$. For weak, but
finite $v$, the level move to the boundaries of the permitted
band:
\begin{eqnarray}\label{en}
E=u\left[\sqrt{1+\Delta^2/u^2}-\frac{v^2}{2}{\left(1+\frac{1}{\sqrt{1+\Delta^2/u^2}}\right)}\right].
\end{eqnarray}

This result can be obtained from the perturbation series on the
alternating signal.

The generalization of Eq. (\ref{en}) for the case of large $v$
reads
\begin{eqnarray}\label{en1}
E=\tilde{u}\left[1+\frac{J_N^2(v)}{2J_N^2(v)-1}\left(1+\sqrt{1+\left[2J_N^2(v)-1\right]\frac{\Delta^2}{\tilde{u}^2}}\right)\right]
\end{eqnarray}

In general, the subsequent local states appear (disappear) when
the energy achieve the permitted band boundaries $|E|=\Delta$.
This gives the threshold equations
\begin{eqnarray}\label{thres}
u=N\pm \Delta[1-J_N^2(v)].
\end{eqnarray}

The numerical results are in accord with the obtained formulae.

\subsection*{How to realize local high-frequency perturbation?}

The most interesting events considered here happen when the
frequency is comparable with the band width and the Fermi energy.
The application of so high frequency is difficult by using of
ordinary electrodes. More natural way is utilization of a freely
propagating electromagnetic wave. (Neglecting the corrections
caused by low-dimensional system itself). The wavelength  of
electromagnetic wave for typical frequency $\omega\gsim E$ exceeds
the electron wavelength, that mean non-locality of electromagnetic
perturbation. Nevertheless, the ways for locality exist. For
example, it can be done using the curved system, by analogy with
the way to produce effectively-nonuniform magnetic field.

\begin{figure}
\centerline{\epsfysize=2.5cm\epsfbox{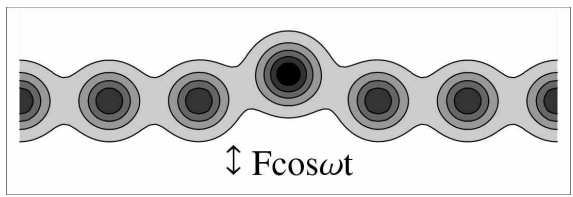}}\caption{Curved
1D quantum dot lattice as an example of considered model. The
potential $U(x,y)=-w\sum_{n\neq 0}\cosh^{-2}\sqrt{(x-n
a)^2+y^2}/l-(w+u)\cosh^{-2}\sqrt{x^2+(y-b)^2}/l$ is supposed. The
Figure represents potential relief for $a=1$, $b=0.3$, $l=0.3$,
$w=1$, $u=0.2$.}\label{relief}
\end{figure}

Let us consider the 1D periodic chain of quantum dots with a
single quantum dot number 0 shifted in $y$ direction in the
distance $b$ (see Fig. \ref{relief}). Being placed into
alternating electric field $F\cos(\omega t)$ in $y$ direction this
system simulates the considered tight-binding model with parameter
$v=F b$. The action of the field on other quantum dots can be
neglected if it is out of resonance with distance between their
levels.

The planar lattice of quantum dots looks quite common, that does
the considered abstract model realistic. Thus, the local probing
of the system by the optical-frequency electric field becomes
possible. The careful examination of this model goes beyond the
topic of the present paper.

In conclusions, we have demonstrated that the local vibrating
potential in 1D system can act as an ideal mirror, despite the
openness of the system. These blockade states co-exist with the
states of ideal transparency and local states.

The work was supported by grant of RFBR No 08-02-00506, the grant
of the President of the Russian Federation No MK-271.2008.2 and
the grant of the Russian Science Support Foundation.

\end{document}